\documentclass[9pt,twocolumn,twoside,lineno]{pnas-new}
\newcommand{\avg}[1]{\langle#1\rangle}
\graphicspath{ {./figures/} }
\usepackage{comment}
\templatetype{pnasresearcharticle} 

\title{Taint analysis of the Bitcoin network
}
\author[a]{Andraž Povše}
\author[a]{Uroš Hercog}

\affil[a]{University of Ljubljana, Faculty of Computer and Information Science, Ljubljana, Slovenia}

\leadauthor{Povše} 

\equalauthors{\textsuperscript{1}A. P. (Andraž Povše) and U. H. (Uroš Hercog) contributed equally to this work.}
\correspondingauthor{\textsuperscript{2}To whom correspondence should be addressed. E-mail: uros@hercog.si (U.H.), ap6932@student.uni-lj.si (A.P)}


\begin{abstract}
\nolinenumbers

Determining the trust of an individual Bitcoin wallet is a difficult problem.
There are no ratings, that offer vendors or exchanges meaningful information about the level of the taint of Bitcoins they are receiving.
Lack of such information places exchanges liable in an event when the received Bitcoins are stolen or ill-gotten.
In this paper, we try to solve this problem by introducing a Bitcoin address taint score called TaintRank. 
It provides insight into a specific wallet by taking the addresses it interacted with throughout history into consideration.
This ranking method provides such Bitcoin exchange companies insight with whom they are trading.
\end{abstract}

\dates{This manuscript was compiled on \today}

\begin{document}
\maketitle
\thispagestyle{firststyle}
\ifthenelse{\boolean{shortarticle}}{\ifthenelse{\boolean{singlecolumn}}{\abscontentformatted}{\abscontent}}{}
\nolinenumbers

\dropcap{C}ryptocurrencies have become a hot topic in the last couple of years with their beginning being contributed to the Bitcoin blockchain created in 2008. 
The project addressed issues regarding money governance, and the issuing being controlled by central authorities. 
It presented a novel solution to these problems using decentralization, consensus protocols such as Proof of Work and promises of privacy. These properties have in the last couple of years attracted a large population of people who saw the potential of such a payments' system being accepted worldwide. This created a need for new financial institutions called cryptocurrency exchanges. These institutions specialized in cryptocurrencies while also providing a bridge to the old system. Most of these institutions lacked the required KYC/AML checks of users which allowed people partaking in illicit activities such as selling drugs and human trafficking to launder their Bitcoins into fiat currencies. 
\section{Motivation}
Another consequence of the quick growth of interest in cryptocurrencies and the lack of proper system security audits were the many hacks of exchanges. 
The hacks dispossessed them of hundreds of thousands of bitcoins. 
Because the Bitcoin network offers pseudo-anonymity to its users, it prevents any linking between addresses and any personally identifiable information\footnote{This assumes that the user is versed in privacy and does not for example reuse addresses. 
It also assumes the use of techniques such as CoinJoin to prevent addresses from being connected. These techniques add additional overhead and are too advanced for the average Bitcoin user so they received little to no traction}. 
Usually, when some bigger hacks happen on popular cryptocurrency exchanges, there is nothing we can do but wait until the thieves want to use their Bitcoins to buy goods or exchange them.
Frequently, users with tainted Bitcoins use mixers to launder them.
A mixer works so that the tainted Bitcoins which are deposited, are put into reserve whereas the clean Bitcoins are paid out to the customer on a new address.
Doing that, we lose the trace of the original address as it can not be directly linked to the new one. \\
This alongside exchanges not verifying their users allows malevolent actors to get by unscathed. With the increasing regulation on cryptocurrencies, financial institutions working with such currencies will be required to do extensive due diligence.
This will be mandatory not only of the involved parties but also of the origin of the coins giving every deposit into their system a risk score. This will raise the awareness about which coins exchanges process to lower their exposure in case of, for example receiving a subpoena from a legal entity investigating, to return them - like credit card charge-backs.
\\
In this paper, we propose a new method of determining the risk factor of received coins from a specific address based on their history and interactions with the network called TaintRank.
By having such information, vendors or exchanges can act differently based on the TaintRank the depositing address has.
Using the TaintRank score, an exchange can require additional verification or a store can refuse participate in the purchase.
Such a ranking system is not expected to be bulletproof, so we should not make harsh actions solely based on TaintRank score a certain address has.

\section{Related work}
The area of cryptocurrency network and transaction analysis has received quite a lot of attention in recent years. Researchers have done extensive research on the analysis of the level of anonymity that the network provides \cite{reidAnonimity} or with the use of different third-party services called mixers or tumblers \cite{de2017analysis,van2018bitcoin,moser2013inquiry}. These services promise to obfuscate the real identities of the owners by randomly exchanging the coins between different users. Such services only disconnect addresses and make direct connections impossible but in reality, they cannot launder the coins in the true essence of the word: someone will always receive marked coins and we can track them even through such services. \\
There has been some research in risk analysis of the Bitcoin network such as in Spagnuolo et al.\cite{spagnuolo2014bitiodine} where they propose a system capable of clustering different addresses, classifying them based on the activity and providing insight into how they interact with the network.
Malovrh \cite{malovrh2018bitcoin} investigates possibilities of using supervised learning as a technique for detecting different anomalies in the Bitcoin network which include criminal activities. 
DebtRank, proposed by Battiston et al. \cite{battiston2012debtrank}, evaluates financial institutions based on the institutions they are loaning or debiting from.
Meiklejohn et al. \cite{meiklejohn2013fistful} deals with understanding the traceability of Bitcoin flows. It gives us a good estimation of how we expect stolen Bitcoins will move and be spent or exchanged. It also touches on some limitations of mixing services, which fail if we are dealing with multiple thousand stolen Bitcoins.

\section{Methods}\label{methods}
Methods we tested are based on node importance. We first construct a directed network where each node represents a Bitcoin address and each directed link a weighted transaction between the two addresses.
Each of the nodes in the network receives its own TaintRank based on different parameters such as node degree and edge weight.
We use prior knowledge to determine the tainted nodes. This knowledge included addresses belonging to a thief who stole Bitcoins from an exchange.
Since the entire Bitcoin network is too large to handle, we use some timeline after a specific event happened. In our case, this was a theft of Bitcoins from an exchange.
We took the data for 1 month after the attack happened and determined the TaintRank score for nodes in the network.\\
We test multiple different approaches to spread the taint throughout the network. First, we test a method where each node spreads its TaintRank by a fixed amount no matter the link weight or other parameters. 
While we would lose track of the specific address that used a mixer, the person behind it will have an equal TaintRank score on their new addresses afterwards.
The negative side of this approach is, that even if an address receives a tiny fraction of Bitcoin from a very tainted address, it itself becomes equally tainted.\\
The further approaches we test take other parameters into the equation.
We use distance from the original thief node, weight of the links, value of the node that is linking to us and other parameters to determine TaintRank of the nodes.
For example, we did not give the same TaintRank score to each of the outgoing transaction, but we set its size proportional to the weight of the transaction.
Doing that, we exclude some nodes not involved in illegal activities but would be tainted using the first method.
But with this, less strict method, we are also open to attacks, that could reduce the TaintRank score of criminal addresses.
\subsection{Hypothesis}
The methods we test differ whether they give out TaintRank proportional to the weight of the link or whether they always give out the same amount no matter the weight.
In the first scenario, we are expecting to be vulnerable to a kind of attack, that spreads Bitcoins into many small fractions, and then ultimately joins some of them together to get almost untainted Bitcoins in some new address.
The second approach would, taint each node that interacts with tainted node (receives Bitcoin from) by a fixed amount.
Because of that, a criminal who knows how the approach works might just send tiny fractions of Bitcoins to random addresses to taint them, with the same amount he is tainted.
Doing that, many regular users will receive a high TaintRank, resulting in more work for the companies trying to filter genuinely tainted addresses.
The perfect combination will most likely comprise some combination of both approaches, taking many parameters into consideration when deciding for the TaintRank score of a node.

\section{Selected data}
Since its inception, the Bitcoin network has grown in size exponentially. It has millions of addresses and hundreds of millions of transaction. This scale poses a great challenge when trying to analyze it. To make the problem more contained, we select a subset of the network which proves to be a good decision. 
We extract 285,591 transactions in a period between 20.06.2011 to 20.7.2011 which amasses to a network with parameters that can be seen in table~\ref{tab:network_data}.

\begin{table}[H]
    \begin{center}
    \caption{Bitcoin network data for the period between 20.06.2011 to 20.7.2011 .}
        \begin{tabular}{ l | c | c }
        \textbf{nodes} & \textbf{links} & \textbf{$\avg{k}$}\\
         \hline
        505,473 & 1,143,444 & 4.524  \\
        \end{tabular}
    \label{tab:network_data}
    \end{center}
\end{table}
The reason we chose this time frame was that one of the bigger Bitcoin thefts happened while the network was still small. This was the Mass MyBitcoin Theft which happened between 20th and 21st of June 2011.
The theft is associated with the address $1MAazCWMydsQB5ynYXqSGQDjNQMN3HFmEu$ which we use as our starting point.
We extract the data using a Bitcoin database index into a public Google BigQuery database which allows us to search with ease for the data using the well-known SQL query language. This removes a lot of issues because we do not have to write a program which fetches transactions from a full-history Bitcoin node which are not public. Using a query in BigQuery we exported all the inputs and the outputs of transactions for that given period. This gives us 100MB JSON file which we further process into a usable graph representation in the edgelist format using a custom script written in Golang. After that, we use scripts written in Python to analyze the graphs and try applying our taint analysis methods.
While processing the data we thought of a heuristic approach which could reduce the size of the overall network. The transactions on the Bitcoin network are made up of different inputs which map to outputs of earlier transactions. To spend an output in a new transaction the owner of the output must provide his signature. Given that a transaction contains inputs from multiple different addresses and that the transaction can be signed only after it was created, we believe it is safe to assume that the one who constructed the transaction is also in control of all the private keys used to sign the transaction. The counter-example to this would be the CoinJoin \cite{maurer2017anonymous} principle. This effectively allows multiple disparate entities to include their unspent outputs into a common transaction. But it also requires a party to act as a synchronizer between all other parties. This creates a high level of complexity which leads us to believe that such principles are rarely used.
Treating addresses of inputs in a transaction as owned by a common entity allowed us to reduce the size of our selected the network by almost a half.

\subsection{Degree distribution}
\begin{figure}[h!]
    \includegraphics[width=8cm]{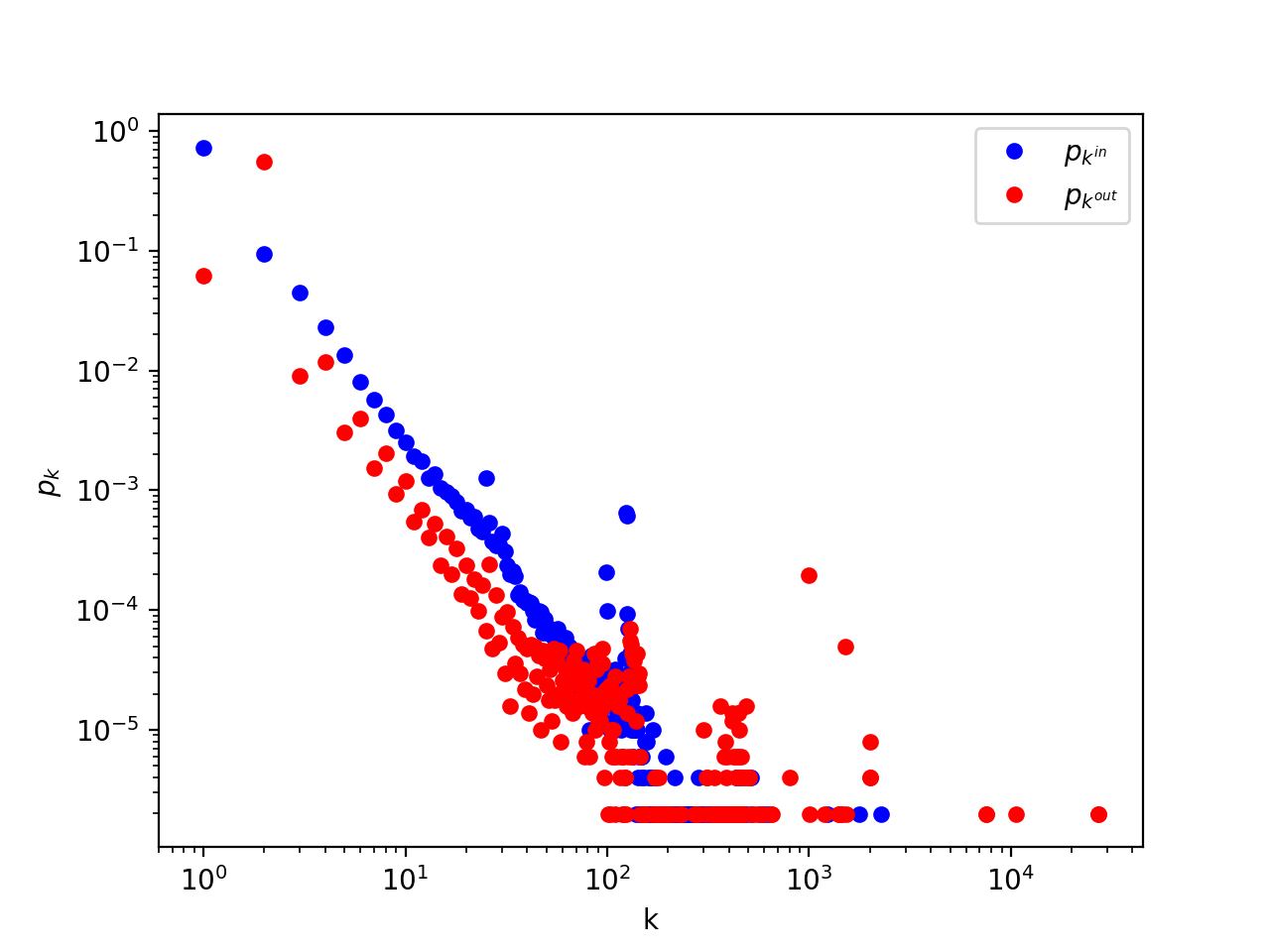}
    \centering
    \caption{Degree distribution in the created network for both in and out degree.}
    \label{fig:degree_dist}
\end{figure}

Degree distribution in our network could tell us a lot about the nodes.
We can assume that nodes with an extremely high degree (hubs) are some sort of exchanges or perhaps mixers.
For example, some nodes are actually Bitcoin exchanges, and we should not spread TaintRank to them since we assume that exchanges would check the incoming addresses, and not accept funds from a (overly) tainted node.
Firstly, we plot the degree distribution for both in and out degrees, which can be seen in figure~\ref{fig:degree_dist}.

As seen from the plot, both vaguely resemble a power-law distribution. 
But considering the spike in the beginning of $p_k^{out}$ and some clearly visible spikes at node degree of 100 and 1000, we should not conclude that the network follows a power-law degree distribution.
For further examination, we look at five addresses with the highest node degree as seen in table~\ref{tab:top_node_deg}

\begin{table}[H]
    \begin{center}
    \caption{Top five Bitcoin addresses in our network that have the highest $p_k^{in}$ and $p_k^{out}$. In the last column there is the believed owner.}
        \begin{tabular}{ p{5cm} | p{0.7cm} | p{1.6cm} }
        \small
        \textbf{Address} & \textbf{Degree} & \textbf{Belongs to}\\
         \hline
         $p_k^{in}$\\
          \hline
          \footnotesize{1eHhgW6vquBYhwMPhQ668HPjxTtpvZGPC} & 2,283  & Mt.Gox\\
          \footnotesize{1PJnjo4n2Rt5jWTUrCRr4inK2XmFPXqFC7} & 1,776  & Slush's pool\\
          \footnotesize{1CxjYUcBFmBftskGFST9MXhyQMQUNTuBzi} & 1,239  & Unknown\\
          \footnotesize{1MAazCWMydsQB5ynYXqSGQDjNQMN3HFmEu} & 627  & MyBTC thief\\
          \footnotesize{1MbSn15MZWNbkNWF72KopyQenCV2zdcWvr} & 598  & Unknown\\
          \hline
         $p_k^{out}$\\
          \hline
          \footnotesize{1MaZAHzEFfinRJ2dwK6YtNDfvWMBkiAxDr} & 27,055  & Unknown\\
          \footnotesize{18qr2srETSvQq4kP7yBYRqQ4LzmjhtRmcD} & 26,975  & Unknown\\
          \footnotesize{1PJnjo4n2Rt5jWTUrCRr4inK2XmFPXqFC7} & 10,607  & Slush's pool\\
          \footnotesize{1AgwESN7RKNZtaqzbqu6kPg3RS6C2qCgHi} & 7,548  & Unknown\\
          \footnotesize{1AZUPm5PC5QguquNsBg7HhWUYz5dfm2nU9} & 7,527  & Unknown\\
          \hline
        \end{tabular}
    \label{tab:top_node_deg}
    \end{center}
\end{table}

We can see that the the highest out-degree is one magnitude higher than the highest in-degree, which tells us that most addresses spread the funds they are sending out to multiple destinations.
Based on the findings of these addresses, we could adjust our algorithms, to take into account what some of these nodes represent.
But the issue is that most of the addresses have an unknown owner.

\section{Results}
With the methods described in ~\ref{methods}, we test our approach on the network.
In the selected data, we know the address where the initial stolen funds went and we base our propagation on it.
Based on that, we start to propagate the taint throughout the network by recursively following all the out links.
The number of nodes which could be reached following the out links from the original address is extremely large (227,656), indicating that the funds have most likely made contact with an exchange of some sort. 
In the table~\ref{tab:subgraph_data}, we can see the characteristics of this subgraph.
When comparing them to the original graph we see that the average node degree is smaller, but no fundamental differences.
\begin{table}[H]
    \begin{center}
    \caption{Subgraph network data for the period between 20.06.2011 to 20.7.2011 .}
        \begin{tabular}{ l | c | c }
        \textbf{nodes} & \textbf{links} & \textbf{$\avg{k}$}\\
         \hline
        227,656 & 379,738 & 3.336
        \end{tabular}
    \label{tab:subgraph_data}
    \end{center}
\end{table}

\begin{figure}[h!]
    \includegraphics[width=8cm]{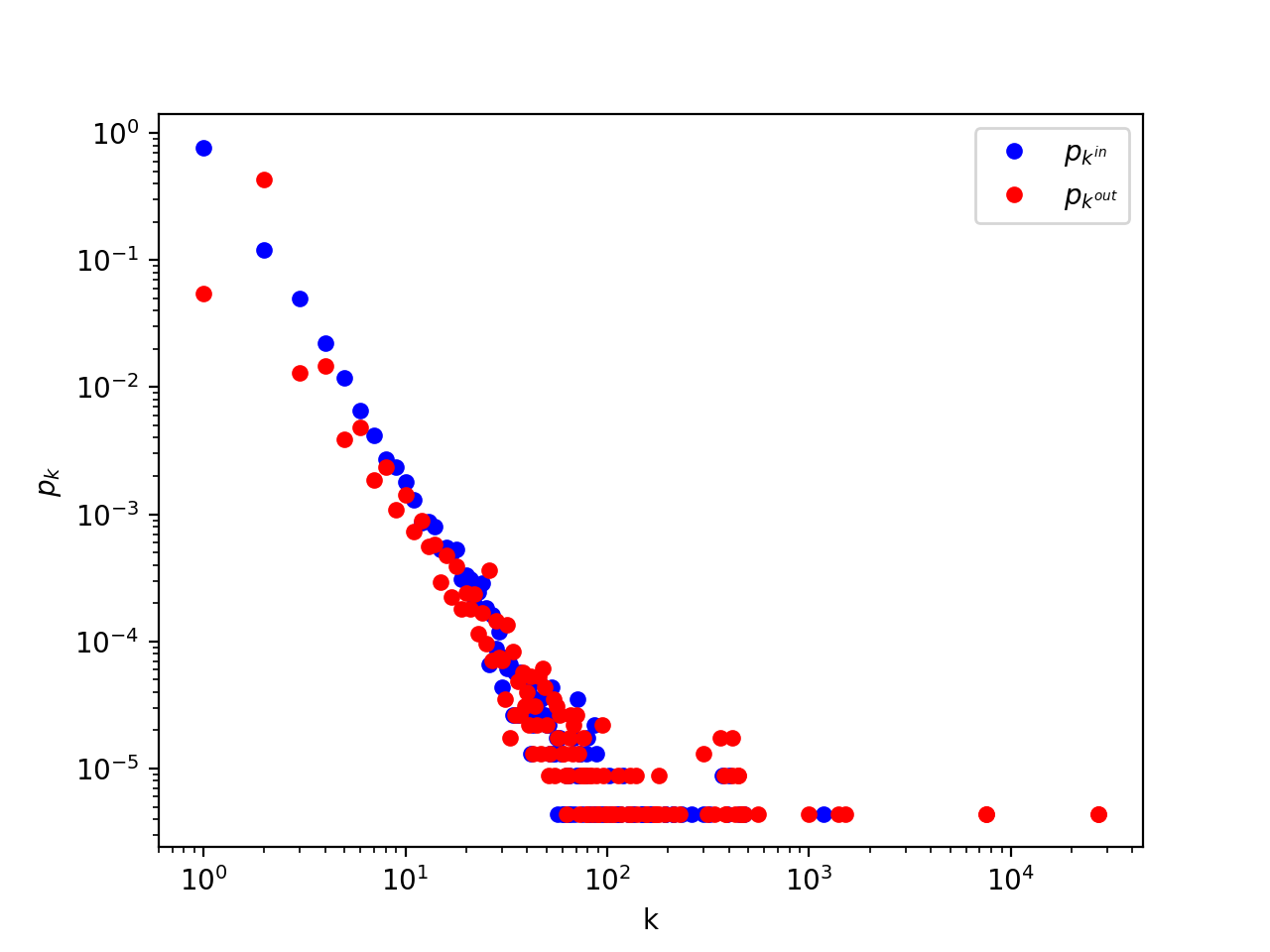}
    \centering
    \caption{Degree distribution in the subgraph created from the network for both in and out degree.}
    \label{fig:subgraph_deg_dist}
\end{figure}

Degree distribution of the subgraph can be seen in figure~\ref{fig:subgraph_deg_dist}.
It seems to follow power-law degree distribution better than the original graph.
The in degree distribution is almost textbook example of power-law, whereas out degree has some spikes that could make us question it.

To reach the last node starting from the tainted one, we have to make 1660 steps.
Based on this observation, we have more than enough nodes, that require different TaintRank scores, since we know they are not equally tainted.
With this, we hope to gain some insight into nodes that are truly tainted, and make sure that nodes that are not really involved with the thief do not get tainted (too much).
In the next approaches, we will be using this subgraph to calculate TaintRank of the nodes.

\subsection{Fixed TaintRank spread}
First method we test consists of applying the TaintRank of the original node to all its successors. 
With this method, as predicted, the drawback was that now we are dealing with almost half the nodes of the original network (227,656) being equally tainted. 
Because of that, we can not differ if some of the nodes are more tainted than the others.
The positive side is that we did not miss any node which received tainted funds, but we also included lots of nodes that should not be tainted.

\subsection{Link weight and node value}
In this method, we spread TaintRank based on the current node value and the weight of the out-link. 
With this approach, we try to take into account the amount of tainted funds that are being sent and how tainted is the address, which is sending us these funds.
We start in the root node and calculate TaintRank for each node in an iterative fashion using the following equation where $V_j$ is the value (sum of all in-links) of node j, $\Gamma_{i}$ represents the in-neighborhood of node $i$ and $w(i,j)$ represents the weight of the edge connecting i and j.

\[t_i = \sum_{j \in \Gamma_{i}} \frac{t_j\times w(j,i)}{V_j}\]

\begin{figure}[H]
    \includegraphics[width=8cm]{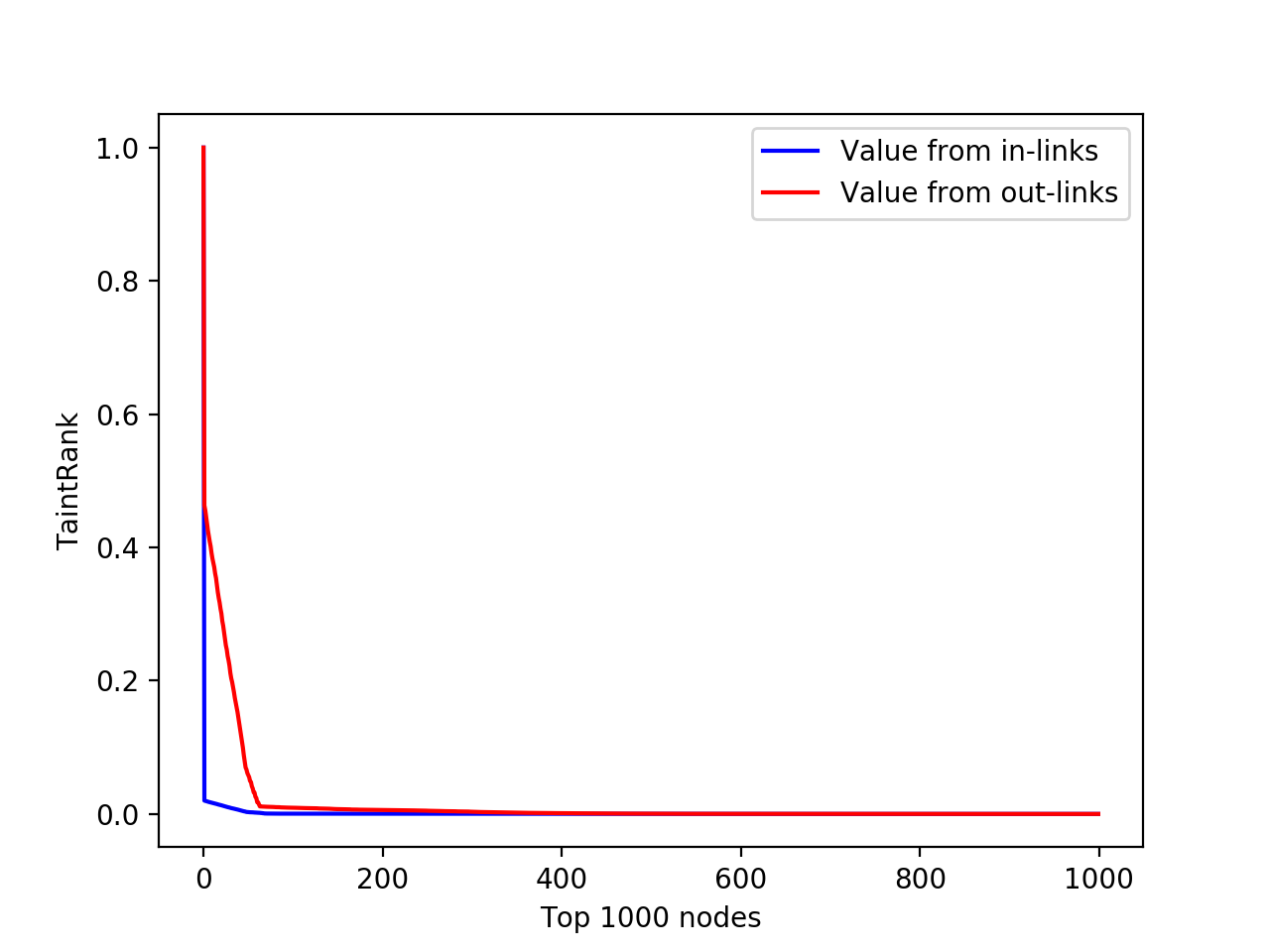}
    \centering
    \caption{Using link weight and node value to determine TaintRank. Displayed are top 1000 highest TaintRank scores for both approaches at determining the value of the node. Y-axis is logarithmic for better separation.}
    \label{fig:link_weight_value}
\end{figure}

Similarly we tested the approach, where we consider $V_j$ as the sum of all out-links instead of in-links.
Reasoning behind it is, that we look only at the funds we are spreading instead of the total funds we are have received.
Following the equation, each node received TaintRank from every one of its predecessors, based on the amount received via this link and the value of the predecessor node.
Results of both approaches can be seen in figure~\ref{fig:link_weight_value}.
As we can see, the approach where we calculated the values from out-links gives a higher TaintRank, since we are only considering spent(sent) funds.
The initial drop is also less drastic than the one with in-links.
With this method, we should consider as tainted roughly first 100 nodes, right until the lines start to have linear pace.

The benefits of this approach are that not all nodes get tainted equally but rather proportional to the amount received.
Negatives include a dropping TaintRank as we split the stolen funds into lots of smaller pieces and spend them separately.

\subsection{Distance from the original tainted node}
In this method we want to take into account distance a certain address has from the root (thief) node.
At start, we set the TaintRank of root node to one and then set the TaintRank for other reachable nodes using iterative fashion.
In the equation we are iterating over all in-links of the currently visited node i and calculating $t_i$ based on their TaintRank and distance of node $i$ from the root node:
\[t_i = \sum_{j \in \Gamma_{i}} \frac{t_j}{distance(i, root)}\]

\begin{figure}[H]
    \includegraphics[width=8cm]{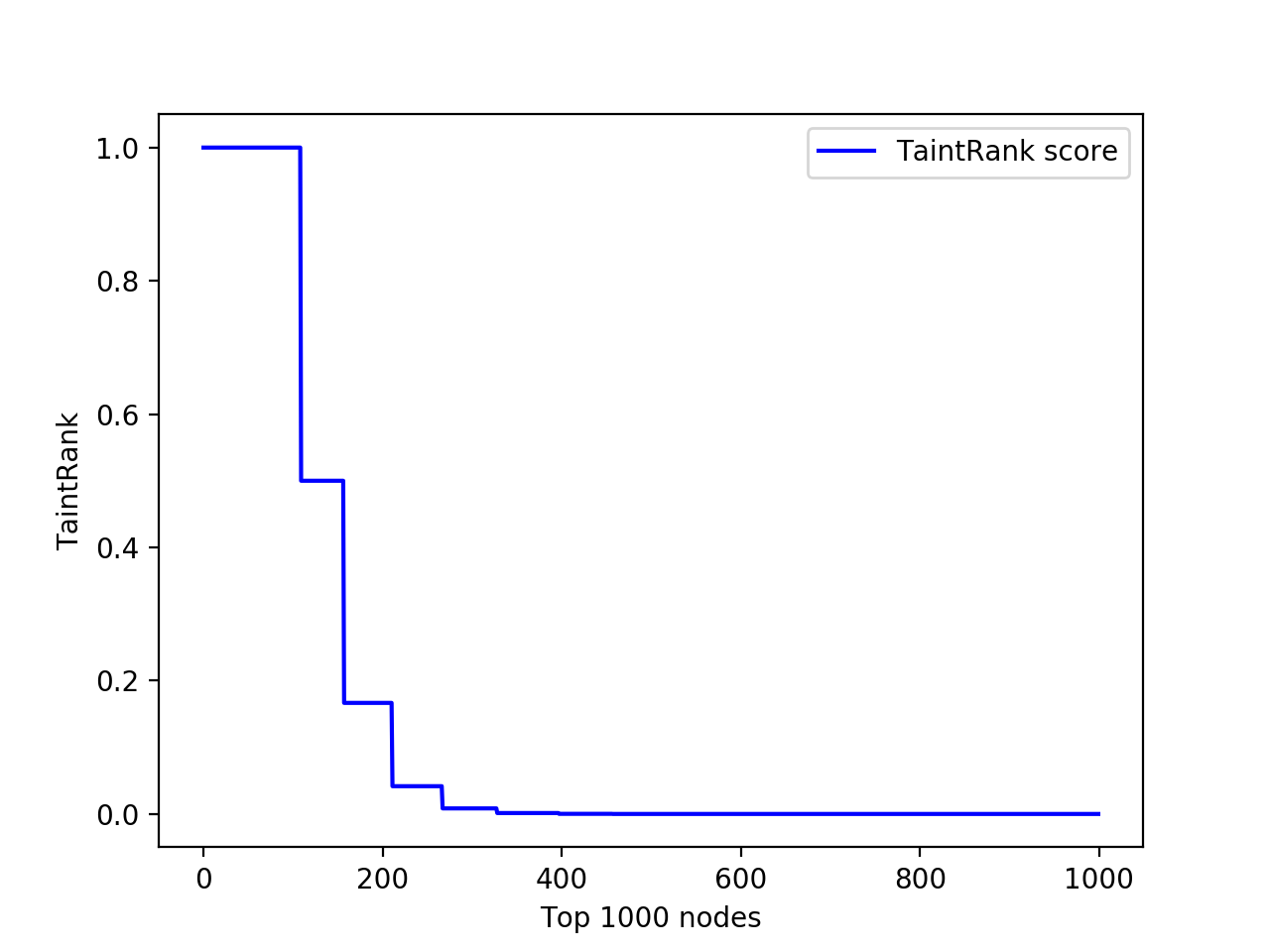}
    \centering
    \caption{Using distance to determine TaintRank. Displayed are top 1000 highest TaintRank scores.}
    \label{fig:link_distance}
\end{figure}

The results for the top 1000 highest TaintRank scores can be seen in figure~\ref{fig:link_distance}.
The top 1000 nodes are not necessarily equal to the ones found in previous method.
With this method, we can capture when a thief is splitting Bitcoins into tiny fractions, but does not make much distance from the original address.
We can see that the curve looks stair-like, where each step represents that we are further from the root node. 
Considering the closeness, we could say that first 100-200 nodes gained relatively large TaintRank and should be considered, whereas the further ones matter less.

Our reasoning for this method was, the further you would be from the tainted node, the less tainted you would get.
Of course this kind of spreading leaves us vulnerable when a thief intentionally makes a long chain before spending the stolen Bitcoins.
\subsection{Combined approach}
With the above, we have seen the shortcomings of each approach.
In order to make it harder for the thief to shake off the taintedness, we try combining both approaches.
Combined approach method calculates taintedness of a node based on both methods. The equation becomes as follows:

\[t_i = \frac{\sum_{j \in \Gamma_{i}} \frac{t_j}{distance(i, root)} + \frac{t_j\times w(j,i)}{V_j}}{2}\]

\begin{figure}[H]
    \includegraphics[width=8cm]{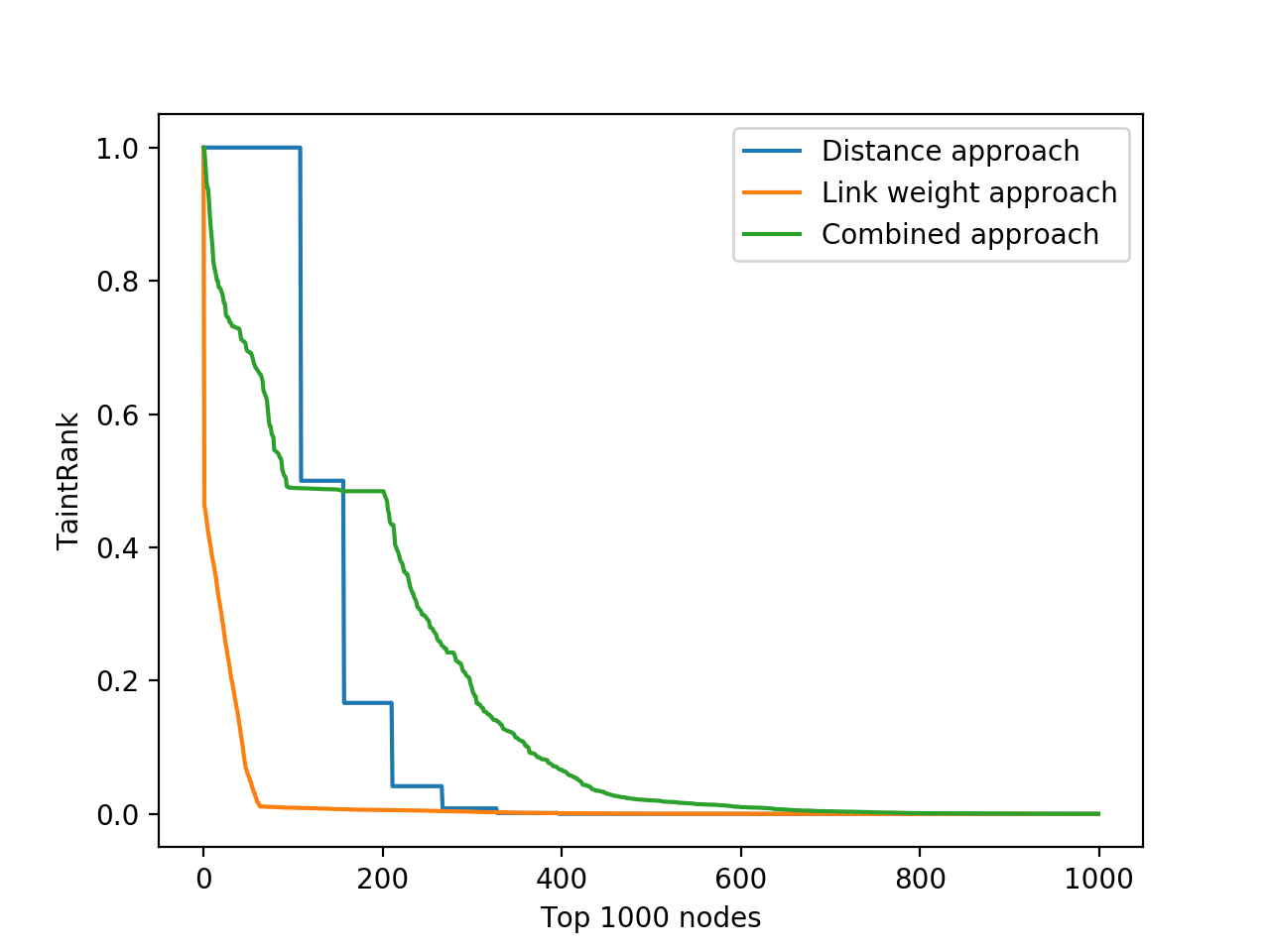}
    \centering
    \caption{Comparison of different approaches at calculating TaintRank. Displayed are top 1000 nodes based on highest TaintRank.}
    \label{fig:all_approaches}
\end{figure}

Of course, because we used both methods, we have to divide final result by two.
We have also tried applying max value from both approaches instead of average between the two, but it produced similar results (around 400 nodes, that would be considered).
In figure~\ref{fig:all_approaches} we can see the differences between the previous approaches and the final, combined one.
On the chart, we can see that the amount of considerably tainted nodes is biggest with the combined approach (around 400 nodes would be considered as tainted).
Reason behind it is that it can handle edge cases, where a specific method would fail.
An example of such edge case, where the weight method would fail is when the thief would spread Bitcoins into tiny fractions across different addresses.
Each would then have a small TaintRank value.
Vice versa can be observed in the distance method, where the thief could make a very long chain of transactions and TaintRank would drop independent of the weight of links.
With this approach, we deal with such cases, where one method fails but the other succeeds.
As with all such problems, even this approach would fail in an event where the thief knows how TaintRank is being calculated and would act to trick it.

\subsection{PageRank like propagation}
For this method we define our taint propagation similarly to how PageRank works. 
First let's define $m_i$ as all in-edges for node $n_i$, $m_i'$ as all tainted in-edges for node $n_i$, and $k_i$ and $k_i'$ as the number tainted and all out-edges for node $n_i$.\\
We define an initial taint value for every node in the network based on the ratio of total tainted in-edges and all in-edges of a node. Labeling of whether an edge is tainted was done during the data preparation phase. 
\[t_i = \frac{m_i'}{m_i}\]
To calculate the actual taint value we use a formula which calculates a nodes taint based on the taints of the neighbour nodes $\Gamma_i$ with whom node $n_i$ has an in-edge
\[t_i = \sum_{e \in \Gamma_i} \frac{t_e}{k_e'}\]
This measure spreads the taint of a node over all its tainted out edges. The distinction of why we use only the tainted out edges is because the node might have legitimate transactions which happened before the ones which moved tainted funds. Such a measure not only marks nodes dealing with a high number of tainted transactions but also propagates this high taint on the nodes that received them. \\
We applied this method on the two instances of both the clustered and non-clustered networks. We also changed the number of times we recalculate the taint on a network by orders of magnitude. A single iteration takes six seconds on a higher range laptop running an eight core i7 processor and 16 GB of RAM. We use a script written in Python to calculate the ranks.\\
\subsubsection{Taint distributions}
\begin{figure}[h!]
\includegraphics[width=8cm]{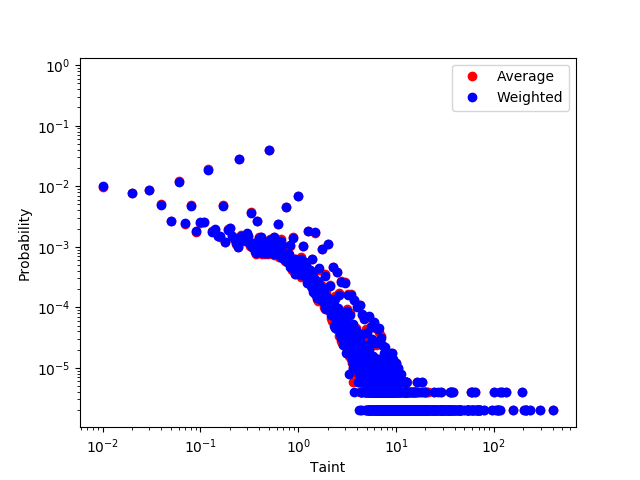}
\centering
\caption{PageRank like propagation for non-clustered with one iteration}
\label{fig:pagerank_1}
\end{figure}

\begin{figure}[h!]
\includegraphics[width=8cm]{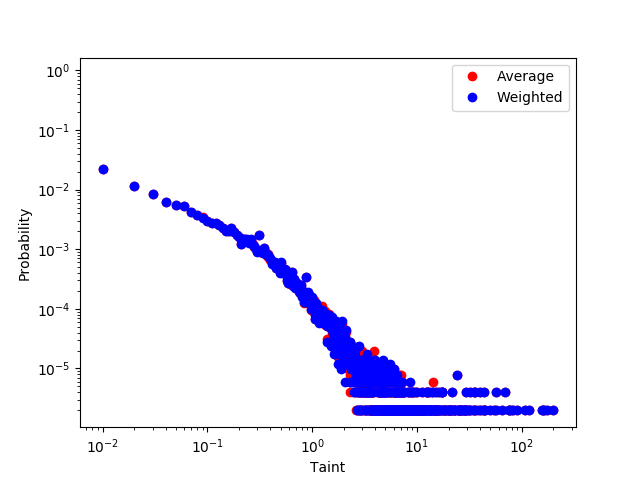}
\centering
\caption{PageRank like propagation for non-clustered with five iterations}
\label{fig:pagerank_5}
\end{figure}

\begin{figure}[h!]
\includegraphics[width=8cm]{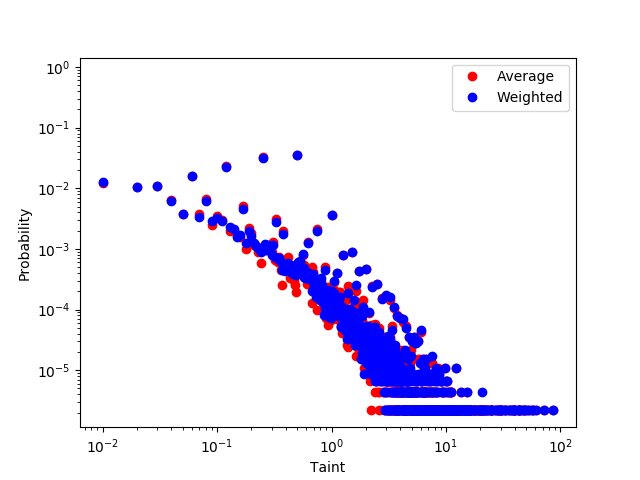}
\centering
\caption{PageRank like propagation for clustered with one iteration}
\label{fig:pagerank_cluster_1}
\end{figure}

\begin{figure}[h!]
\includegraphics[width=8cm]{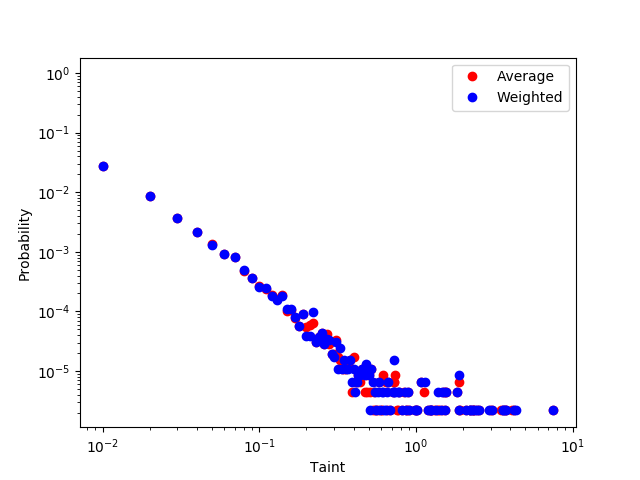}
\centering
\caption{PageRank like propagation for cluster with five iterations}
\label{fig:pagerank_cluster_5}
\end{figure}

We run the method on a non-clustered and clustered graph with one and five iterations. After calculating the values, we plot the taint distributions of these graphs. Figures \ref{fig:pagerank_1} and \ref{fig:pagerank_5} represent the distributions for non-clustered graph, while \ref{fig:pagerank_cluster_1} and \ref{fig:pagerank_cluster_5} represent the clustered graph. What we can observe from these figures is that with the increase of iterations, the taint distribution seems to resemble power-law distribution. Such a distribution is the most expressed with the clustered five iterations graph taint distribution. \\
The reason for such a distribution is difficult for us to reason about and it requires further investigation. It might be a coincidence, based on the data sample we worked on or it might have some to us unknown underlying mechanics.

\subsubsection{Nodes with the highest taint rank}
We select the best five nodes for every iteration count according to their taint rank. We then try searching for these addresses on the internet to try and reason on why they have such high ranks. This is easy for the non-clustered graph because every node is its own address. But with the clustered graph this proves to be a challenge. Not only do we have to check all the addresses in a cluster, but also because we have to reason why the clustering method works.\\
The top node with the highest taint rank in non-clustered graph with one iteration is $1eHhgW6vquBYhwMPhQ668HPjxTtpvZGPC$. After performing search on the internet, we believe address belongs to an exchange. Search on the internet lead us to believe that the address is connected to an exchange called Mt. Gox. Other four addresses seem to be burner addresses with high amounts of incoming and outgoing bitcoins. They could potentially be used by one of the mixing services.\\
Searching for the top nodes with the highest taint rank in non-clustered graph with five iterations gave us no information about the addresses' main purpose. The top address from the single iteration result was not on the list anymore while the other four addresses moved up a spot. A new address took the fifth place. Most of the addresses again look like burner addresses. To determine whether this is correct, we'll have to find a way to classify them.



%
%
%

\section{Discussion}
We've developed five different methods and all of them give at least slightly different results. This makes the evaluation even more difficult. Knowing whether a specific method's results are relevant, without understanding the underlying socioeconomic mechanics, makes this problem even more difficult to tackle.
Results provided by the approaches make a lot of sense.
Since we were calculating TaintRank on 45 \% nodes of the entire network, they must not be all highely tainted - the vast majority should be very mildly tainted.
This was achieved using all of the approaches.
We have seen the pros and cons of each and tried some combinations of them.
Entire computations are computationally very efficient with time complexity for the first four iterative approaches of $O(n)$.
The assigned scores give us some insight on what nodes we would like to avoid as a Bitcoin exchange company or a merchant.
If we would process these ill-gotten Bitcoins as an exchange company or vendor we might face some liabilities.
Potentially we would have to refund the person that was hacked and deal with the loss ourselves.
Such company could also be treated as a suspect in the theft.
As stated before, we should not solely rely on this rank, but it can be a first step we take.
Potential use case for the TaintRank system would be performing such scores in real time.
After an exchange got hacked, they would pinpoint the original hacker node and TaintRank would be calculated for the future nodes, preventing others from accepting funds from tainted addresses.
Same could be applied to some nodes, that are known to participate in illicit activities.
In a perfect scenario, we could rely solely on the most basic method we tested.
Giving each node that is reachable from the thief node a TaintRank of one and assume complete knowledge from all traders - so legitimate people would not accept funds from the tainted person.
But this is not feasible, so other approaches we tested would probably be applied.

\section{Conclusion}
Analyzing the Bitcoin network is a big challenge just by itself.
The ever-expanding nature makes it very difficult to process the whole network fast enough for it to stay relevant with every new block. 
With the increasing popularity the number of unique addresses and transactions between them grow exponentially. 
But given enough resources and sufficiently powerful processing infrastructure, this isn't something we couldn't solve. The bigger issue is the social, non-technical aspect of the network. 
The questions of how to validate the results of developed methods. Different methods return different results and comparing them without any prior knowledge is next to impossible.
We believe that the methods we developed are a great step in the right direction. They show us that a taint score can be calculated for addresses. Given enough context, it can prove to be useful in determining whether an address is trustworthy. But we also believe that these methods are far from being applicable in real world scenarios where the livelihood of a crypto-currency exchange is at stake. 

\section{References}
\bibliography{bibliography}

\begin{thebibliography}{1}

\bibitem{reidAnonimity}
{Reid} F, {Harrigan} M (2011) {An Analysis of Anonymity in the Bitcoin System}.
\newblock {\em arXiv e-prints} p. arXiv:1107.4524.

\bibitem{de2017analysis}
de~Balthasar T, Hernandez-Castro J (2017) An analysis of bitcoin laundry
  services in {\em Nordic Conference on Secure IT Systems}.
\newblock (Springer), pp. 297--312.

\bibitem{van2018bitcoin}
van Wegberg R, Oerlemans JJ, van Deventer O (2018) Bitcoin money laundering:
  mixed results? an explorative study on money laundering of cybercrime
  proceeds using bitcoin.
\newblock {\em Journal of Financial Crime} 25(2):419--435.

\bibitem{moser2013inquiry}
M{\"o}ser M, B{\"o}hme R, Breuker D (2013) An inquiry into money laundering
  tools in the bitcoin ecosystem in {\em 2013 APWG eCrime Researchers Summit}.
\newblock (Ieee), pp. 1--14.

\bibitem{spagnuolo2014bitiodine}
Spagnuolo M, Maggi F, Zanero S (2014) Bitiodine: Extracting intelligence from
  the bitcoin network in {\em International Conference on Financial
  Cryptography and Data Security}.
\newblock (Springer), pp. 457--468.

\bibitem{malovrh2018bitcoin}
Malovrh J (2018) Bitcoin anomalies analysis.

\bibitem{battiston2012debtrank}
Battiston S, Puliga M, Kaushik R, Tasca P, Caldarelli G (2012) Debtrank: Too
  central to fail? financial networks, the fed and systemic risk.
\newblock {\em Scientific reports} 2:541.

\bibitem{meiklejohn2013fistful}
Meiklejohn S, et~al. (2013) A fistful of bitcoins: characterizing payments
  among men with no names in {\em Proceedings of the 2013 conference on
  Internet measurement conference}.
\newblock (ACM), pp. 127--140.

\bibitem{maurer2017anonymous}
Maurer FK, Neudecker T, Florian M (2017) Anonymous coinjoin transactions with
  arbitrary values in {\em 2017 IEEE Trustcom/BigDataSE/ICESS}.
\newblock (IEEE), pp. 522--529.

\end{thebibliography}

\showacknow{} 
\end{document}